\documentclass[12pt,aps,preprint,a4paper,showpacs]{revtex4}
\usepackage{amsmath}
\usepackage{graphicx}

\newcommand{\re}{\ref}
\newcommand{\be}{\begin{equation}}
\newcommand{\ee}{\end{equation}}

\newcommand{\la}{\label}
\newcommand{\ber}{\begin{eqnarray}}
\newcommand{\eer}{\end{eqnarray}}

\begin{document}
\title{A SMALL PARAMETER APPROACH FOR FEW--BODY PROBLEMS}

\author{V.D. Efros\footnote{E-mail efros@mbslab.kiae.ru}}

\affiliation{ RRC "Kurchatov Institute"}


\begin{abstract}
A procedure to solve few--body problems is developed which is based on an expansion over a small 
parameter. The parameter is the ratio of potential energy to kinetic energy 
for states having not small hyperspherical quantum numbers, 
$K>K_0$.
Dynamic equations are reduced perturbatively to equations in the finite--dimension subspace with 
$K\le K_0$. Contributions from states with $K>K_0$ are taken into account in a 
closed form, i.e. without an expansion over basis functions. Estimates on efficiency of 
the approach are presented.
\end{abstract}

\bigskip 


\bigskip 

\maketitle

\section{Introduction}

While the nuclear force cannot be treated as a perturbation, the
potential energy of, say, a bound state of a nucleus is in fact comparable to its kinetic energy only for a limited number of 
its components.
These are components having low values of the hypermomentum $K$.
For all remaining components kinetic energy exceeds potential energy  which allows
solving few--nucleon problems perturbatively. 

This approach is
suggested below and is a development of that of Ref. \cite{2} where a 
perturbation method
has been given to solve large systems of bound--state linear equations pertaining to the hyperspherical--hyperradial 
expansion.\footnote{In an early paper \cite{3}
the wave function component with $K=2$ was calculated perturbatively from that with $K=0$. However,
the change of the component with $K=0$ itself due to coupling to $K=2$ has not been taken into
account. This missed quantity is in general of the same magnitude as that accounted for.} The method  
proved to be efficient 
\cite{4,5}. The expansion
parameter was the  potential--to--kinetic--energy matrix element ratio. However, at A$>$3 it is the calculation of matrix elements
themselves  that requires a massive computational effort. 

The difficulty stems from a swift rise, as $K$ increases,
of the number of hyperspherical harmonics having the same $K$. (The larger is a number of particles
the swifter is the rise.) Selection of hyperspherical  harmonics to reduce the computational effort, see \cite{2,6,7}, is not
efficient for A$>$4 bound states and is not justified in reaction 
calculations.  In the method given below the difficulty is removed. No expansions are employed
when $K$ is not small.

Recently a considerable progress in methods for solving few--body problems has been achieved.
However, those developments have limitations not arising in the presented method.
In particular, the well--known Green Function Monte Carlo (GFMC) method 
is the method to calculate a bound state of a system, and it does not suit to calculate
reactions. (Although the simplest scattering problems may be considered in its frames.) 
Unlike this method, the method presented below is
suitable for calculating reactions of a general type. 
Besides,
the GFMC method is not convenient in that providing separate observables 
it does not provide the wave function of a bound
state as an outcome of the calculation. And in the framework of the method described below 
bound state wave functions are calculated in a rather simple form suitable for subsequent applications. 

Recently a way was found to extend the Faddeev--Yakubovsky A=4 calculations over the energy range above 
the four--body breakup threshold \cite{8}. However,  Yakubovsky type calculations require too much 
numerical effort. The amount of calculations is less in the scheme 
below.

At solving few--body problems, convergence of results for calculated 
observables was accelerated with the help of the effective interaction methods.
Such  methods were developed in the framework of the oscillator expansion  \cite{9} and
the hyperspherical expansion \cite{10}. In this approach, a true Hamiltonian is replaced with
some effective Hamiltonian acting in a subspace of only low excitations. When, formally, the
latter subspace is enlarged an effective Hamiltonian
turns to a true one. An effective Hamiltonian is constructed from a requirement that its ingredients, 
as defined in the subspace of low excitations, reproduce some properties of the corresponding
ingredients of a true Hamiltonian in the total space. It has been shown \cite{9,10} that this, indeed, 
leads to improvement of convergence for observables considered.

Higher excitations are disregarded in such type calculations. It is clear, however, that correlation
effects related to higher excitations cannot be reproduced by any state vector lying in the allowed 
subspace of only low excitations. Consider e.g. the mean value, 
$\langle\Psi_0|H|\Psi_0\rangle$, of such an "observable" as a true Hamiltonian. 
It follows from the variational principle that an approximate state $\Psi_0$ obtained with such 
a method provides poorer approximation to the true  $\langle\Psi_0|H|\Psi_0\rangle$ value than 
$\Psi_0$ given by the simple diagonalization of a Hamiltonian in the same subspace of low 
excitations. But even the latter $\langle\Psi_0|H|\Psi_0\rangle$ value 
is a very poor approximation for realistic  Hamiltonians. On the contrary, the method given
below provides an approximate state vector  that is apparently close to a true state vector both 
as to its low excitation component and to its high excitation component.

And speaking of reaction calculations in the framework of the dynamic schemes employed below 
one should also take into account that a rate of convergence is
determined not only by the properties of a  Hamiltonian but also by those of the source--term $q$
entering the equations.
But these properties are apparently ignored at constructing effective  Hamiltonians. 
Unlike this, the method developed below provides
state vectors genuinely close to the true ones both for bound state problems and reaction problems.

Development of efficient microscopic methods for nuclear physics is timely now because of   
a necessity to test nuclear forces derived from the effective field theory. 

The method is given in the next section, and its implementation is considered  in Sec. 3. In Sec. 4.
estimates on its efficiency are presented. In particular, the method seems to be promising to test
realistic nuclear forces  
  in four--nucleon reaction problems and in scattering problems below three--body breakup thresholds in the 
  $^5$He, $^7$Li, $^7$Be, and $^8$Be systems.
 This opens up basically a new field.

\section{The small parameter expansion}

We consider first the bound state problem
\be
(H-E_i)\Psi_i =0,\la{1}
\ee
where $H=T+V$ is an A--body Hamiltonian. We 
split the whole space of states into the subspaces with $K\le K_0$ and $K>K_0$ and we denote 
$\Psi_i^l$ and $\Psi_i^h$ the components of the solution $\Psi_i$ that lie, 
respectively, in these subspaces.  
Let us denote $P_{K_0}$ and $Q_{K_0}$ projectors onto
the $K\le K_0$ and $K>K_0$ subspaces, respectively. Let  $E_i^{(0)}$ 
be the zero approximation
eigenvalue that arises when coupling to the $K>K_0$ subspace is disregarded,
\be
P_{K_0}\left(H-E_i^{(0)}\right)\Psi_i^{l(0)} =0.\la{2}
\ee
The state $\Psi_i^{l(0)}$ is the corresponding  zero approximation
eigenfunction. 

Let us also define a Green function in the $K>K_0$ subspace.
The operators of hyperrotations commute with the kinetic energy operator. Therefore,   
if $\Psi_Q$ is a state lying in the $K>K_0$ subspace then $T\Psi_Q$ also belongs to this 
subspace. We shall consider also the "correction operators" $\delta H$ specified below 
that have the same propety. Then 
\be\varphi_Q=(T+\delta H -E_i^{(0)})\Psi_Q\la{137}\ee 
is a state 
belonging to the  $K>K_0$ subspace. We also assume that $T+\delta H$ possess only continuum spectrum.
Then for any state $f_Q$ in the $K>K_0$ subspace one has \mbox{$(T+\delta H -E_i^{(0)})f_Q\ne0$}. Taking 
also into account that the operator from (\re{137}) possess a complete set of eigenstates with $K>K_0$
one concludes that   
there exists a unique state $\Psi_Q$ corresponding to any given $\varphi_Q$ in (\re{137}). Thus 
we can define the corresponding Green function 
\be
G_{K_0}=\left(T+\delta H-E_i^{(0)}\right)^{-1}Q_{K_0}.\la{3}
\ee

Taking into account that 
$P_{K_0}TQ_{K_0}=Q_{K_0}TP_{K_0}=0$ we write down   Eq. (\re{1}) in the form
\ber
P_{K_0}(H-E_i)\Psi_i^{l} =-P_{K_0}V\Psi_i^{h},\la{4}\\
\Psi_i^h=-G_{K_0}\left[V\Psi_i^{l}+\left(U-\Delta E_i\right)
\Psi_i^h\right],\la{5}
\eer
where \[U=V-\delta H,\qquad \Delta E_i=E_i-E_i^{(0)}.\]
When the hyperangular momentum $K_0$ is sufficiently large kinetic energies of states in the
$K>K_0$ subspace are high. Then in accordance with Eq. (\re{3}) $\Psi_i^h$ is "small", so that 
one may treat the second term in the right--hand side of Eq. (\re{5}) as a perturbation. Thus
one may express the component $\Psi_i^h$ in terms of $\Psi_i^l$
perturbatively and obtain dynamic equations for the latter component alone. 
Namely, one may write
\be 
\Psi_i^h=-\Gamma V\Psi_i^{l} ,\la{6}
\ee
where 
\be 
\Gamma=G_{K_0}-G_{K_0}(U-\Delta E_i)G_{K_0}
+G_{K_0}(U-\Delta E_i)G_{K_0}(U-\Delta E_i)G_{K_0}-\ldots.\la{7}
\ee
(We mention that, say, the contribution from $\Delta E_i G_{K_0}^2$ here is not of the 
same magnitude as that from $G_{K_0}UG_{K_0}$.)
The dynamic equation (\re{1}) may then be put in the form of an equation in the subspace of low excitations only,
\be 
P_{K_0}(T+V-V\Gamma V-E_i)\Psi_i^{l} =0.\la{8}   
\ee
The quantity $-V\Gamma V$ represents a genuine effective interaction that reproduces the effect of the excluded
$K>K_0$ subspace. 

It is convenient to treat perturbatively not only subsequent contributions to the $\Gamma$ operator but also
the whole $-V\Gamma V$ interaction. This leads to  an expansion  over powers of both
$G_{K_0}U$ and $G_{K_0}V$.
We replace $G_{K_0}$ with $\lambda G_{K_0}$ and we seek for $\Psi_i^l(\lambda)$ as an expansion,
\be
\Psi_i^l(\lambda)=\sum_{m\ge0}\lambda^m\Psi_i^{l(m)},\qquad E_i(\lambda)=\sum_{m\ge0}\lambda^mE_i^{(m)}\la{9}
\ee
setting $\lambda=1$ at the end. The zero order eigenstates and eigenvalues are given by Eq. (\re{2}). Equations for higher
order corrections are of the form
\be
P_{K_0}(H-E_i^{(0)})\Psi_i^{l(m)}=q^{(m)}\la{10}
\ee
with a given source term. Explicit form of the equations  for $\Psi_i^{l(1)}$ and
$\Psi_i^{l(2)}$ is 
\be
P_{K_0}(H-E_i^{(0)})\Psi_i^{l(1)}=E_i^{(1)}\Psi_i^{l(0)}+
P_{K_0}VG_{K_0}V\Psi_i^{l(0)},\la{11}
\ee
\ber 
P_{K_0}(H-E_i^{(0)})\Psi_i^{l(2)}=E_i^{(1)}\Psi_i^{l(1)}+
E_i^{(2)}\Psi_i^{l(0)}
+P_{K_0}VG_{K_0}V\Psi_i^{l(1)}
\nonumber\\
-P_{K_0}VG_{K_0}UG_{K_0}V\Psi_i^{l(0)}.\la{12}
\eer
(The term from (\re{7}) with $\Delta E_i\rightarrow E_i^{(1)}$ would first appear  at $m=3$.)

Obviously, solutions to Eqs. (\re{10}) are not unique: at given lower order corrections entering
$q^{(m)}$ the solution $\Psi_i^{l(m)}$ is determined  up to const$\cdot\Psi_i^{l(0)}$. 
It is convenient to select unique solutions imposing e.g. the usual perturbation theory condition
\be\langle\Psi_i^{l(0)}|\Psi_i^{l}(\lambda)\rangle=\langle\Psi_i^{l(0)}|\Psi_i^{l(0)}\rangle.\la{13}\ee 
This is a normalization condition since up to a numerical
factor $\Psi_i^{l}(\lambda)$ are determined by Eq. (\re{1}) type equations.
According to the first of
Eqs. (\re{9}) this is equivalent to the set of conditions
\be
\langle\Psi_i^{l(0)}|\Psi_i^{l(m)}\rangle=0,\qquad m=1,2,\ldots.\la{14}
\ee 
(Use of any other condition of Eq. (\re{13}) type whose right--hand side includes
additional terms tending to zero at $\lambda\rightarrow 0$ would lead merely to a different
normalization of  the sum of $\Psi_i^{l(m)}$.) 

In order Eqs. (\re{10}) be self--consistent the conditions  $\langle\Psi_i^{l(0)}|q^{(m)}\rangle=0$ should be fulfilled.
This determines the energy corrections $E_i^{(m)}$.  ( In the present consideration complications that arise in cases of 
possible non--trivial (quasi) degeneracies of levels are disregarded.) 
We get, in particular,
\be
E_i^{(1)}=-\frac{\langle\Psi_i^{l(0)}|VG_{K_0}V|\Psi_i^{l(0)}\rangle}
{\langle\Psi_i^{l(0)}|\Psi_i^{l(0)}\rangle},\la{15}
\ee 
\be
E_i^{(2)}=\frac{\langle\Psi_i^{l(0)}|VG_{K_0}UG_{K_0}V|
\Psi_i^{l(0)}\rangle-\langle\Psi_i^{l(0)}|VG_{K_0}V|\Psi_i^{l(1)}\rangle}
{\langle\Psi_i^{l(0)}|\Psi_i^{l(0)}\rangle}.\la{16}
\ee
Since by construction $G_{K_0}$ is positive definite the correction $E_i^{(1)}$ is negative. (Besides,
one has in (\re{16}) $\langle\Psi_i^{l(0)}|VG_{K_0}V|\Psi_i^{l(1)}\rangle=-\langle\Psi_i^{l(1)}|H-E_i^{(0)}|\Psi_i^{l(1)}\rangle.$)

Thus we have dynamic equations of the Eqs. (\re{9}), (\re{10}) form  which do not involve excitations with $K>K_0$
at all.   This is an advantage
since the $K\le K_0$ subspace spans only a finite, hopefully not too large, number of HH. 

The complementary $K>K_0$ component of a state sought for is given by Eq. (\re{6}). This can be represented as
\be
\Psi_i^{h}=\sum_{m\ge1} \Psi_i^{h(m)},\la{17}
\ee
where, in particular,
\be
\Psi_i^{h(1)}=-G_{K_0}V\Psi_i^{l(0)},\la{18}
\ee
\be
\Psi_i^{h(2)}=-G_{K_0}V\Psi_i^{l(1)}+G_{K_0}U
G_{K_0}V\Psi_i^{l(0)}.\la{19}
\ee

The low excitation component $\Psi_i^{l}$ is obtained above in the form of a hyperspherical expansion.
This component may be stored in this form for use in the relations of the Eqs. (\re{18}), (\re{19}) type and also in other applications. 
The complementary high--excitation component 
$\Psi_i^{h}$  is reconstructed as a quadrature and may be applied in such a form. This
component does not involve any expansion. 

According to the variational principle the energy eigenvalue calculated as the average value of a Hamiltonian 
over $\Psi_i^{l(0)}+\Psi_i^{l(1)}+\Psi_i^{h(1)}$
is accurate up to the order $m=3$.

Let us also comment on applying of a such type expansion for calculation of reactions. First we shall proceed in the framework
of the following well--known procedure \cite{11,12}. Consider reactions below the
three--fragment breakup threshold  when only two--fragment channels are open. Let $\Psi_i$ be a  continuum spectrum state
and denote $N$ the number of open two--fragment channels. 
The ansatz
\be
\Psi_i=\phi_i^{(1)}+\sum_{j=1}^Nf_{ij}\phi_j^{(2)}+X,\la{20}
\ee
is used where $\phi_i^{(1)}$ and $\phi_j^{(2)}$ represent the "channel" states of two possible types, 
$f_{ij}$ are reaction amplitudes to be determined, while $X$ 
is localized and is sought for as an expansion over hyperspherical harmonics. In the three--nucleon case the procedure
is applicable also above the three--nucleon breakup threshold.  In this case $X$ describes breakup to free particles 
at large distances. The equation determining $X$ is
\ber
(H-E)X=q,\la{21}\\
q=-{\bar \phi}_i^{(1)}+\sum_{j=1}^Nf_{ij}{\bar \phi}_j^{(2)},\nonumber
\eer 
where ${\bar \phi}_i^{(1),(2)}=(H-E)\phi_i^{(1),(2)}$ are localized states. The state $X$ is found
from Eq. (\re{21}) up to reaction amplitudes $f_{ij}$ (cf. below), 
\be
X=X_i+\sum_{j=1}^Nf_{ij}X_{j},\la{22}
\ee and 
$f_{ij}$ are obtained from $N$ additional linear equations. 

Proceeding as above we represent $X$ as $X^l+X^h$. We have
\be
P_{K_0}\left(T+V-V\Gamma V-E\right)X^l=P_{K_0}\left(q-V\Gamma q\right),\la{23}
\ee
\be
X^h=-\Gamma\left(VX^l-q\right).\la{24}
\ee
Here $\Gamma$ is of the form of Eq. (\re{7}) with $\Delta E_i$ being omitted
and with $E_i\rightarrow E$ in Eq. (\re{3}). Writing as above
$X^l=\sum_mX^{l(m)}$,  $X^h=\sum_mX^{h(m)}$ one obtains equations for $X^{l(m)}$
and expressions for $X^{h(m)}$ performing an expansion over $G_{K_0}$. 
One has, in particular,  
\be
P_{K_0}(H-E)X^{l(0)}=P_{K_0}q\la{25}
\ee
while the corresponding equations for $X^{l(1),(2)}$ and 
expressions for $X^{h(1),(2)}$ are obtained from Eqs. (\re{11}),  (\re{12}) and (\re{18}),  (\re{19})
with $E_i^{(1),(2)}$ being omitted and with the replacements $E_i^{(0)}\rightarrow E$, and
$\Psi^{l,h(1),(2)}\rightarrow X^{l,h(1),(2)}$, $V\Psi^{l(0)}\rightarrow VX^{l(0)}-q$.

One may note that Eq. (\re{21}) has a localized solution only when
the amplitudes $f_{ij}$ equal to their true values. Otherwise, the solution includes an admixture
of cluster components. On the contrary, if a finite number of terms is retained in the expansion
of $\Gamma$ over powers of $G_{K_0}$ then the localized solution exists at any  $f_{ij}$ allowing
the representation (\re{22}). The reason is that the role of higher terms in the expansion of  $\Gamma$
increases at large distances. When  $f_{ij}$ in (\re{21}) not coincide with their true values
these higher terms are responsible for description of the large distance clusterization (cf. below).
The same occurs in the approximate way to solve Eq. (\re{21}) applied up to now when the
equation of the form (\re{25}) was employed. Of course, this does not pose any problems. 
 
In the general type reaction case the method is applicable in the framework of the approach \cite{13,14,15}
in which reaction observables are obtained from states $\tilde\Psi$ that vanish at large distances like bound states.  
The approach
extensively applied for perturbation induced reactions and proved to be very efficient. 
Any strong--interaction induced reactions can also be treated in this way.  
Dynamic equations for states  $\tilde\Psi$ are of the form of Eq. (\re{21}) with $X\rightarrow \tilde\Psi$ 
where $q$ is a given state and the energy $E$ is complex.
Because of the latter the solution $\tilde\Psi$ is localized and it
is a proper object to be found with the help of the expansion over $G_{K_0}$.
The perturbative expansion to calculate $\tilde\Psi=\tilde\Psi^l+\tilde\Psi^h$ is similar to that described above. 

The outcome of a calculation are quantities of the form 
$\Phi(E)=\langle{\tilde\Psi}'(E)|{\tilde\Psi}(E)\rangle$
where $\tilde\Psi'$ is a state similar to  $\tilde\Psi$ for another source term $q'$.
Reaction observables are extracted from $\Phi(E)$  in a simple way as quadratures. 
When  it is sufficient to calculate $\Phi(E)$ only up to the first
order in $G_{K_0}$  
one need not account for ${\tilde\Psi}^{h}$ whose contribution in $\Phi(E)$ is of the second order.  
    
\section{Implementation}

We are dealing with the space of Jacobi coordinates or that of Jacobi momenta and we denote $n=3A-3$ the space dimension. 
Considering matrix elements in the momentum representation we denote $\bar\pi$ the
$n$--dimensional momentum vectors 
and $\Pi=|{\bar\pi}|$ the hypermomentum. We
adopt in (\re{7}) $\delta H={\bar V}(\Pi)$ where ${\bar V}(\Pi)$ is a subsidiary interaction.
We have 
\be
\langle{\bar\pi}'|G_{K_0}|{\bar\pi}\rangle=\frac{\delta^{(n)}({\bar\pi}'-{\bar\pi})-[\delta(\Pi'-\Pi)/\Pi^{n-1}]
\sum_{K\le K_0;\nu}Y_{K\nu}^*({\hat \pi}')Y_{K\nu}({\hat \pi})}
{\Pi^2/(2m)+{\bar V}(\Pi)-E},\la{26}
\ee
\be
\langle{\bar\pi}'|U|{\bar\pi}\rangle=\langle{\bar\pi}'|V|{\bar\pi}\rangle
 -\delta^{(n)}({\bar\pi}'-{\bar\pi}){\bar V}(\Pi),\la{27}
\ee
where for bound states $E$ means $E_i^{(0)}$. Here $Y_{K\nu}$ is an orthonormalized complete
set of hyperspherical harmonics with a given $K$,
and ${\hat \pi}$ denotes a unit vector pointed in the direction of  $\bar \pi$, 
${\hat \pi}={\bar\pi}/\Pi$, ${\hat \pi}'={\bar\pi}'/\Pi'$.
The hyperangular factor entering (\re{26}) may be 
represented with the simple expression (e.g. \cite{16})
\be
\sum_\nu Y_{K\nu}^*({\hat \pi}')Y_{K\nu}({\hat \pi})=\frac{K+\frac{n-2}{2}}{2\cdot\pi^{n/2}}
\Gamma\left(\frac{n-2}{2}\right)C_K^{\frac{n-2}{2}}({\hat \pi}'\cdot{\hat \pi}),\la{28}
\ee
where $C_K^\gamma(x)$ is the Gegenbauer polynomial.

When performing calculations in the coordinate representation we denote $\xi$ the $n$--dimensional 
position vectors, $\rho=|\xi|$, and $\hat\xi=\xi/\rho$. We have
\be
T=T_\rho+\frac{\hbar^2}{2M}\frac{{\hat K}^2}{\rho^2},\la{29}
\ee
\be
\langle\xi'|T_\rho|\xi\rangle=\delta^{(n)}(\xi'-\xi)\left(-\frac{\hbar^2}{2M}\right)
\left(\frac{d^2}{d\rho^2}+\frac{n-1}{\rho}\frac{d}{d\rho}\right).\la{30}
\ee
Here ${\hat K}^2$ is the hyperangular momentum operator.  In this case we choose:
\be
\delta H={\bar V}(\rho)-T_\rho+E,\la{31}
\ee
where ${\bar V}(\rho)$ is a subsidiary interaction. 
It  is convenient to represent the corresponding $G_{K_0}$ as a sum of contributions from various $K$ values,
\be
G_{K_0}=\sum_{K>K_0}g_K.\la{32}
\ee
Then
\be
\langle\xi'|g_K|\xi\rangle=\left[\frac{\hbar^2}{2M}\frac{K(K+n-2)}{\rho^2}+{\bar V}(\rho)\right]^{-1}
\frac{\delta(\rho'-\rho)}{\rho^{n-1}}\frac{K+\frac{n-2}{2}}{2\cdot\pi^{n/2}}
\Gamma\left(\frac{n-2}{2}\right)C_K^{\frac{n-2}{2}}({\hat \xi}'\cdot{\hat \xi}),
\la{33}\ee
\be
\langle\xi'|U|\xi\rangle=\langle\xi'|V|\xi\rangle-\delta^{(n)}(\xi'-\xi)\left[\bar{V}(\rho)-T_\rho+E\right].
\la{34}\ee

The choice (\re{31}) is done to facilitate Monte--Carlo calculations of matrix elements. 
In the  coordinate representation, $G_{K_0}$ that contains the total $T$ as in Eq. (\re{26}) would correspond to 
$g_K$ with hyperradial Green functions varying rapidly at not small $K$.

When performing calculations it is 
convenient to include 
the factor $\sum_\mu|\theta_\mu\rangle\langle\theta_\mu|\equiv I$ in the Green functions, 
where $\{\theta_\mu\}$ is a complete set of spin--isospin states  (c.f. \cite{4}) with simple permutational properties. 

The  subsidiary interactions ${\bar V(\rho)}$ and ${\bar V(\Pi)}$ can be chosen from the requirement of
fast convergence of observables as  $K_0$ increases when one takes into account only, say, the
lowest order corrections.  Alternatively, one can minimize the ratio  of the  second order
correction to the first order correction for this purpose.

If e.g. in the coordinate representation one writes (suppressing the spin--isospin notation)
\be
\Psi_i^{l(m)}=\sum_{K\le K_0;\nu}\chi_{K\nu}^{(m)}(\rho)Y_{K\nu}({\hat \xi})\la{35}
\ee
then the equations of Eqs. (\re{10}) -- (\re{12}) and (\re{25}) type are of the form 
\be
-\frac{\hbar^2}{2M}\left(\frac{d^2}{d\rho^2}+\frac{n-1}{\rho}\frac{d}{d\rho}-\frac{K(K+n-2)}{\rho^2}\right)  
\chi_{K\nu}^{(m)}-E\chi_{K\nu}^{(m)}+\sum_{K'\nu'}(K\nu|V|K'\nu')\chi_{K'\nu'}^{(m)}=(K\nu|q),\la{36}
\ee
where $K\le K_0$ and  $K'\le K_0$. 

Their right--hand sides as well as the $\Psi^{h(m)}$
components are to be calculated with the Monte-Carlo method. 
Integrands depend on high $K$ values only via Gegenbauer polynomials (\re{28}) entering (\re{26}) and (\re{33}).
While these polynomials are rather quickly oscillating all other factors in the integrands are smooth functions of coordinates or momenta.
In the case $m=1$ one may simplify a calculation taking the argument of Gegenbauer polynomials 
as a new integration variable. Integration over this variable may be done with the help of the regular Gauss--Gegenbauer
quadratures while
integration over other variables that are smooth may be done with the Monte--Carlo method. A suitable change of variables
is described in Appendix. This can also be done in the case of a momentum representation calculation.
At the same time there are indications (e.g. \cite{17})
that direct
Monte--Carlo integration may be suitable even at  rather large $K$ values.

Local components are dominating components of nuclear forces derived from the effective field theory.
If $m\ge2$ correction terms are retained in a calculation a reasonable simplification may be to account for only those local 
components in these terms.  

Direct solution of Eqs. (\re{36}) in the form they are written down is hampered
by the large centrifugal barriers $K(K+n-2)/\rho^2$. 
In the bound--state case a practical procedure is to expand $\chi_{K\nu}^{(m)}(\rho)$ 
over a set of functions that reduces Eqs. (\re{36}) to linear equations. 
Such linear equation sets of a large size may efficiently be solved with a version of the method of Ref. \cite{2},
i.e. using an expansion over another parameter of the $K_0^{-2}$ type. 
For complex $E$ values with positive real parts and rather small imaginary parts entering Eqs. (\re{36}) in the reaction calculations convergence
of the expansion procedure to solve Eqs. (\re{36}) is slow. And for real $E$ values entering  Eqs. (\re{36}) in the other type calculations
of reactions this procedure 
may lead to unphysical
singularities in reaction observables. Other efficient solution methods are available for this purpose. 

Integral transforms $\Phi(E)$ are required at sufficiently
many values of complex energies $E$ to perform a satisfactory inversion \cite{13,14,15}. But one need 
not solve Eqs. (\re{36}) for all these $E$ values. A better approach is to solve these 
equations 
for a rather  small number of $E$ values and to obtain $\Phi(E)$ for a larger set of $E$
values via interpolation. The transforms $\Phi(E)$ are smooth functions and this procedure is
safe and accurate.

\section{Estimates}

First let us consider numerical estimates for the $^4$He system in the 0$^+$ state.  In \cite{4} 
the accurate $\alpha$--particle binding energy pertaining  to a proper subset
of hyperspherical harmonics has been compared with energies calculated approximately as follows. 
Only matrix elements of NN force $(K\nu|V|K'\nu')$ such that either $K$ or $K'$ does not exceed some $K_0$ were retained
in the system of equations. This approximation is equivalent to the approximation $\Gamma\rightarrow G_{K_0}$ in Eq. (\re{8}).
The approximate solution thus obtained is close to that given by the $m=1$ approximation of the method described above. It accounts
for corrections to binding energy up to $m=3$. The corresponding
 approximate energy values as a function of $K_0$  along with the accurate value are shown in
Table 4 of Ref.  \cite{4}. 

At $K_0=14$ the difference between the two values equals to 0.23 MeV for 
the NN interaction with a very strong repulsive core. In this connection one needs to remember that the binding energy
considered is a small difference between two large quantities, potential and kinetic energy, which
deteriorates the accuracy.  This comparison refers to the case when the above mentioned mean field ${\bar V}$ 
is set to zero. Inclusion of the mean field in the calculation would improve the convergence. Furthermore, it is seen from  Table 4 of   \cite{4}
that for another NN force having  a softer core convergence with respect to $K_0$ is much faster. One may note in this connection that  
nuclear forces derived from the effective field theory are much softer than that used in the above comparison. For such forces convergence 
with respect to a maximum $K$ value retained in a calculation is
considerably faster than for phenomenological local realistic forces \cite{18}. Naturally, for effective field theory forces one may  expect 
faster
convergence as to $K_0$ as well. 

Now let us perform an estimate for the case of  reactions in the same system. We consider the above outlined 
approach dealing with complex energy $E$, and we set $E=\sigma_R+i\sigma_I$. We take $\sigma_I=10$ MeV which
is a good value for e.g. electromagnetic processes \cite{15}. We consider again the case $K_0=14$ and again we shall not
include the mean field ${\bar V}$. We employ the so called AV4 potential that is the central component of the 
realistic AV18 NN interaction \cite{19}. For the estimate purposes we adopt the following model. We represent the subspace with $K\le K_0$
 with a single hyperspherical harmonic with $K=0$. We represent the subspace with $K>K_0$   with a single 
hyperspherical space--symmetric "potential" harmonic with $K=16$. Thus in our model we deal with two coupled differential 
equations that have the form  of
Eqs. (\re{36}) and that correspond to the hyperspherical expansion of the state $\tilde \Psi$ described in the preceding section. 
The right--hand side source term in the first of the equations was set to be $\exp(-\rho/0.4\,{\rm fm})$. The form of the
source terms is not very important for the estimate and in the second of the equations the source term  was set to zero. In general, 
at large $\rho$ values the
state $\tilde \Psi$ contains the 3N+N cluster components, in particular. When $K$ is less or about $\rho/\sqrt{3}R$, 
$R$ being the range of  the 3N cluster, in equations of the type we consider matrix elements may produce coherent effects.
Then it would not be suitable for our estimates to represent the $K>K_0$ subspace with a single hyperspherical harmonics.
In our case,  however, a typical extension in $\rho$ of the state $\tilde \Psi$ is 10 fm, c.f. below, 
and with our $K_0$ value these effects are not relevant. The results for the quantity of interest $\Phi(E)=\langle\tilde \Psi(E)|\tilde \Psi(E)\rangle$
 are shown in Fig. 1. The curves labeled as the first and the second approximation correspond, respectively,  to calculations 
in the frameworks of $m=1$ and $m=2$ approximations as described in the preceding section. We also note that the central force we use for the estimate
has a strong repulsive core of the height of 2.7 GeV. As above, one may note that use of a nuclear force  derived from the effective field theory
and inclusion of the mean field  ${\bar V}$ in a calculation would improve the rate of convergence.  
\begin{figure}
\begin{center}
\includegraphics[scale=0.5, angle=-90]{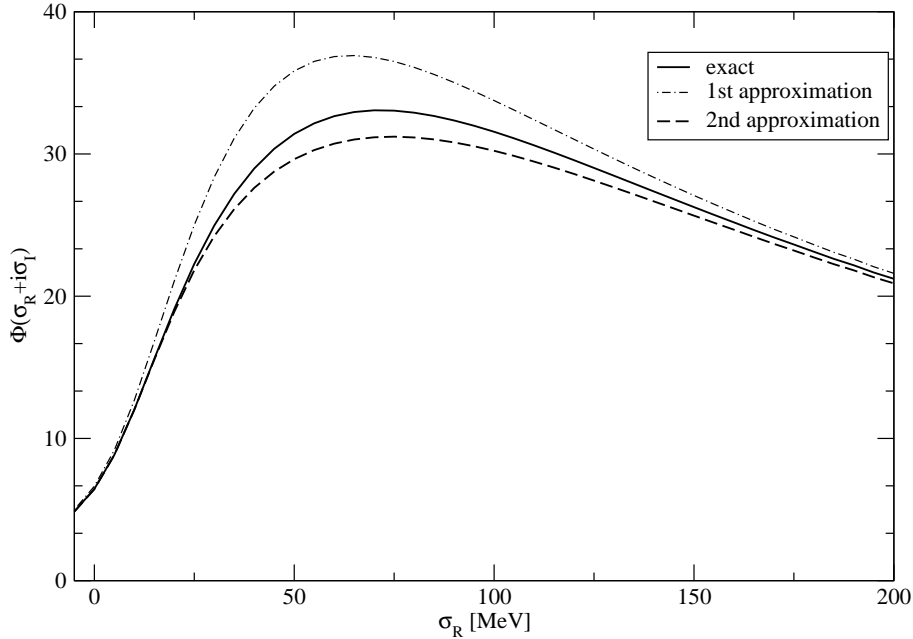}
\caption{Integral transform of the form factor of the type of those governing reaction
amplitudes or representing response functions. The calculation is done for the model
described in the text. The quantity $\sigma_I$ was set to be 10 MeV.  Exact values and the m=1 and m=2
approximations are compared.  }
\end{center}
\end{figure}

In the problems we considered there exists several hundreds hyperspherical 
harmonics with $K\le 14$ that is acceptable. A swift rise of the number of hyperspherical harmonics with the same $K$ starts at $K$ values
about 10, see Table 3 from \cite{4}. So it would be desirable to keep $K_0$ at the level of 10--12 in a calculation. 

In general,  conditions for applicability of the perturbation approach we consider are
\be
||G_{K_0}U\Psi^h||\ll||\Psi^h||,\qquad  ||G_{K_0}V\Psi^l||\ll||\Psi^l||.\la{37}
\ee
A better accuracy in the first of these conditions is provided reducing $U$ via inclusion of the mean field ${\bar V}$.
The second of these conditions contains the interaction in the form $Q_{K_0}VP_{K_0}$. It is non--diagonal in hyperspherical
quantum numbers and is substantially "smaller" than $Q_{K_0}VQ_{K_0}$ for this reason. For bound states and for scattering states 
in the most interesting case when 
${\rm Re} E$ is smaller or about $|\langle V-{\bar V}\rangle|$ 
the first
of these conditions, for example,  may roughly be represented as
\be
\frac{\hbar^2}{2m{\bar \rho}^2}\left(K_0+\frac{n-2}{2}\right)^2\gg|\langle V-{\bar V}\rangle|,\la{38}
\ee 
where ${\bar \rho}\simeq A^{1/2}\langle r^2\rangle^{1/2}$ is the corresponding range  
either of the bound state $\Psi$, or that of the component $X$ of the continuum state above, or that of the state
$\tilde\Psi$ above while $\langle V-{\bar V}\rangle$ is the average value over $\Psi$, or $X$, or $\tilde\Psi$.\footnote{If one wants
to derive Eq. (\re{38}) with use of the momentum representation one needs to take into account that 
if a coordinate representation wave function $f(\xi)$ is localized within a
hyperradius $\rho$ then the momentum representation quantities $(Y_{K\nu}({\hat \pi})|f(\pi))$
are very small at $\Pi$ values such that $\Pi\rho\ll K+(n-2)/2$.} 
In the case, for example, of the $^7$Li two--cluster  ground state the rate of convergence may be increased if one diminishes ${\bar \rho}$
separating out the cluster 
component of the state (c.f. Eq. (\re{20}). 

The value of $\bar \rho$ in (\re{38}) in the case of  the $\tilde\Psi$ state is about $({\rm Im} k)^{-1}$ where $(\hbar k)^2/(2M)=E$. If 
$E=\sigma_R+i\sigma_I$ then ${\rm Im} k=2^{-1/2}[(\sigma_R^2+\sigma_I^2)^{1/2}-\sigma_R]^{1/2}$. The choice of $\sigma_I$ is discussed 
in \cite{15}, see also \cite{20}.

It occurs that when ${\rm Re} E=\sigma_R$ increases at given $K_0$ the left--hand sides in Eqs. (\re{37}) decrease if $G_{K_0}$ depends on $E$
as in Eq. (\re{26}).

We note that the separation out of two--cluster components to diminish the ${\bar \rho}$ value may also be done when calculating  
the $\tilde\Psi$ quantities. 
 At large distances the relative motion factor  is an outgoing (neutral or Coulomb) wave  with
a complex wave number having a positive imaginary part in this case. This ensures vanishing of these components at large 
distances.
    
In the n+n+p continuum--state case the method may be applied also above the three--body breakup threshold 
despite the fact that the component $X$ is not localized. The reason is that average potential energy for the total breakup channel decreases 
with $\rho$ as $\rho^{-3}$. But in the 3--body case it may be profitable to do the whole calculation in the matrix form i.e. to use
the HH expansion for the perturbative calculation also of the $K>K_0$ contributions.
 
In conclusion, an area for testing realistic nuclear forces may be substantially extended with the help of the  presented approach.  
For this purpose, it is required to investigate the issue of Monte--Carlo computing the perturbative corrections. 

The work was partially supported by RFFI, grant 07-02-01222a, and RMES, grant NS-3004.2008.2.
 
\bigskip

\appendix*

\section{}

When one takes ${\hat \xi}'\cdot{\hat \xi}$ as a new integration variable one needs 
to define the whole set of integration variables in a way that the integrand remains non--singular. This can 
be done e.g. as follows. Let us express a unit vector ${\hat \xi}=\{{\hat \xi}_1,\ldots,{\hat \xi}_n\}$ in 
terms of another unit vector ${\hat \eta}$, 
\[{\hat \xi}_i=\sum_{j=1}^ng_{ij}{\hat \eta}_j,\]
where $g_{ij}$ is an orthogonal matrix such that its first column is $g_{i1}={\hat \xi}_i'$ and
$g_{ij}$ is arbitrary otherwise. One then has 
${\hat \xi}'\cdot{\hat \xi}=\sum_{i,j}g_{i1}g_{ij}{\hat \eta}_j={\hat \eta}_1$. Let us parametrize the 
components of ${\hat \eta}$ as follows,
\[{\hat \eta}_1=\cos\varphi,\qquad  {\hat \eta}_j={\hat v}_{j-1}\sin\varphi,\qquad j=2,\ldots,n,\]
where ${\hat v}_{i}$ are components of a unit vector ${\hat v}$ on a hypersphere in a $n-1$--dimensional
subspace. Taking into account that
\[ d{\hat \xi}=d{\hat \eta}\equiv(\sin\varphi)^{n-2}d{\hat v}d\varphi\]
one then may rewrite e.g. integrals of the structure $\langle F_1|G_{K_0}|F_2\rangle$  as
\ber
 \sum_{K>K_0} \frac{K+\frac{n-2}{2}}{2\cdot\pi^{n/2}}\Gamma\left(\frac{n-2}{2}\right)
 \int \rho^{n-1}d\rho (\sin\varphi)^{n-2}d\varphi d{\hat v} d{\hat \xi}'
F_1^*(\rho{\hat \xi})\nonumber\\
\left[\frac{\hbar^2}{2m}\frac{K(K+n-2)}{\rho^2}+{\bar V}(\rho)\right]^{-1}
C_K^{\frac{n-2}{2}}(\cos\varphi)F_2(\rho{\hat \xi}'),\nonumber
\eer 
where the components of the $n$--dimensional unit vector ${\hat \xi}$ entering $F_1$ are
parametrized as follows,
\[{\hat \xi}_i={\hat \xi}_i'\cos\varphi+
\left(\sum_{j=2}^ng_{ij}({\hat \xi}'){\hat v}_{j-1}\right)\sin\varphi.\]
The integrations over $d\rho$, $d{\hat \xi}'$, and $d{\hat v}$ may be performed with the Monte--Carlo
method while the remaining integration over $d\varphi$ may be done with the help of regular 
quadratures. Integrals at a given $\rho$ over the hypersphere which pertain to the $m=1$ correction are transformed similarly.

\end{document}